\documentclass[12pt,twoside]{article}
\usepackage{fleqn,espcrc1}

\usepackage{graphicx}
\usepackage[figuresright]{rotating}

\newcommand{\AmS}{{\protect\the\textfont2
  A\kern-.1667em\lower.5ex\hbox{M}\kern-.125emS}}

\title{Orbitally Excited Baryons in Large $N_c$ QCD\thanks{Invited parallel
session talk presented at PANIC99, Uppsala Sweden, June 10, 1999.
William and Mary preprint no. WM-99-113.}}
\author{Christopher D. Carone
\address{Nuclear and Particle Theory Group \\
College of William and Mary, Williamsburg, VA 23187 USA} 
        \thanks{Research supported in part by the National Science Foundation
under Grant Nos.\ PHY-9800741 and PHY-9900657.}
}
       
\begin{document}

\maketitle

\begin{abstract}
We present a model-independent analysis of the mass spectrum of nonstrange 
$\ell=1$ baryons in large $N_c$ QCD. The $1/N_c$ expansion is used to 
select and order a basis of effective operators that spans the nine 
observables (seven masses and two mixing angles). Comparison to the data 
provides support for the validity of the $1/N_c$ expansion, but also reveals
that only a few nontrivial operators are strongly preferred.  We show that 
our results have a consistent interpretation in a constituent quark 
model with pseudoscalar meson exchange interactions.
\end{abstract}

\section{Introduction} \label{sec:intro}

        QCD admits a useful and elegant expansion in powers of $1/N_c$, 
where $N_c$ is the number of colors~\cite{tHooft}.  This expansion has 
been utilized successfully in baryon effective field theories to isolate 
the leading and subleading contributions to a variety of physical 
observables~\cite{revs}.  

Here we study the mass spectrum of the nonstrange, $\ell=1$ baryons 
(associated with the SU(6) {\bf 70}-plet for $N_c=3$) in a large-$N_c$ 
effective theory~\cite{CCGL1,CCGL2}.   We describe the states as a 
symmetrized ``core'' of $(N_c-1)$ quarks in the ground state plus one 
excited quark in a relative $P$ state.  ``Quarks'' in the effective theory 
refer to eigenstates of the spin-flavor-orbit group, SU(6) $\times$ O(3),
such that an appropriately symmetrized collection of $N_c$ of them have 
the quantum numbers of the physical baryons.  Baryon wave functions 
are antisymmetric in color and symmetric in the spin-flavor-orbit indices 
of the quark fields.  While this construction assures that we obtain 
states with the correct total quantum numbers, we do not assume that 
SU(6) is an approximate symmetry of the effective Lagrangian.  Rather, 
we parameterize the most general way in which spin and flavor symmetries 
are broken by introducing a complete set of quark operators that act on 
the baryon states.  Matrix elements of these operators are hierarchical 
in $1/N_c$, so that predictivity can be obtained without recourse to 
ad hoc phenomenological assumptions. 

The nonstrange {\bf 70}-plet states which we consider in this analysis
consist of two isospin-$3/2$ states, $\Delta_{1/2}$ and $\Delta_{3/2}$, 
and five isospin-$1/2$ states, $N_{1/2}$, $N^\prime_{1/2}$, $N_{3/2}$, 
$N^\prime_{3/2}$, and $N^\prime_{5/2}$.  The subscript indicates total
baryon spin; unprimed states states have quark spin $1/2$ and primed 
states have quark spin $3/2$.  These quantum numbers imply
that two mixing angles, $\theta_{N1}$ and $\theta_{N3}$, are necessary 
to specify the total angular momentum $1/2$ and $3/2$ nucleon mass 
eigenstates, respectively.  Definitions for these angles can be found 
in Ref.~\cite{CGKM}.

\section{Operators Analysis} \label{sec:melements}

To parameterize the complete breaking of SU(6)$\times$O(3), it is 
natural to write all possible mass operators in terms of the generators 
of this group.   The generators of orbital angular momentum are denoted
by $\ell^i$, while $S^i$, $T^a$, and  $G^{ia}$ represent the spin,
flavor, and spin-flavor generators of SU(6), respectively. The generators 
$S_c^i$, $T_c^a$, $G_c^{ia}$ refer to those acting upon the $N_c-1$ core 
quarks, while separate SU(6) generators $s^i$, $t^a$, and $g^{ia}$
act only on the single excited quark.  Factors of $N_c$ originate
either as coefficients of operators in the Hamiltonian, or through
matrix elements of those operators.  An $n$-body operator, 
which acts on $n$ quarks in a baryon state, has a coefficient of 
order $1/N_c^{n-1}$, reflecting the minimum number of
gluon exchanges required to generate the operator in QCD.
Compensating factors of $N_c$ arise in matrix elements if sums over 
quark lines are coherent. For example, the unit operator $1$ contributes 
at $O(N_c^1)$, since each core quark contributes equally in the matrix 
element. The core spin of the baryon $S_c^2$ contributes to the masses 
at $O(1/N_c)$, because the matrix elements of $S_c^i$ are of $O(N_c^0)$ 
for baryons that have spins of order unity as $N_c \to \infty$.  Similarly, 
matrix elements of $T_c^a$ are $O(N_c^0)$ in the two-flavor case since 
the baryons considered have isospin of $O(N_c^0)$, but the operator 
$G_c^{ia}$ has matrix elements on this subset of states of $O(N_c^1)$.  
This means that the contributions of the $O(N_c)$ quarks add incoherently 
in matrix elements of the operator $S_c^i$ or $T_c^a$ but coherently 
for $G_c^{ia}$. Thus, the full large $N_c$ counting of the matrix element
is $O(N_c^{1-n+m})$, where $m$ is the number of coherent core quark 
generators. A complete operator basis for the nonstrange {\bf 70}-plet masses
is shown in Table~1\footnote{Some of these operators were considered
previously in Ref.~\cite{goity}.}. Index contractions are left implicit 
wherever they are unambiguous, and the $c_i$ are operator coefficients.  The 
tensor $\ell^{(2)}_{ij}$ represents the rank two tensor combination 
of $\ell^i$ and $\ell^j$ given by 
$\ell^{(2)}_{ij} = \frac{1}{2} \{ \ell_i , \ell_j \} - \frac{\ell^2}{3}
\delta_{ij}$.
\begin{table}[htb]
\caption{A complete operator basis, ${\cal O}_i$, $i=1\ldots 9$, for the 
nonstrange {\bf 70}-plet masses.}
\label{matel}
\newcommand{\m}{\hphantom{$-$}}
\newcommand{\cc}[1]{\multicolumn{1}{c}{#1}}
\renewcommand{\tabcolsep}{2pc} 
\renewcommand{\arraystretch}{1.2} 
\begin{tabular}{@{}lllll}
\hline
$c_1${\bf 1} & $c_2\ell s$  & $c_3\frac{1}{N_c} \ell^{(2)} gG_c$ & & \\
$c_4(\ell s + \frac{4}{N_c+1} \ell tG_c)$ & $c_5\frac{1}{N_c} \ell S_c$ &
$c_6\frac{1}{N_c} S_c^2 $ & & \\
$c_7\frac{1}{N_c} tT_c $ & $c_8\frac{1}{N_c}\ell^{(2)} sS_c $ 
& $c_9\frac{1}{N_c^2} \ell^i g^{ja} \{ S_c^j,G_c^{ia} \}$ & & \\ 
\hline
\end{tabular}\\[2pt]
Operators $1$, $2$--$3$, and
$4$--$9$ have matrix elements  of order $N^1_c$,
$N^0_c$, and $N^{-1}_c$, respectively.
\end{table}
\section{Results} \label{sec:results}

        Since the operator basis in Table~1 completely spans
the $9$-dimensional space of observables, we can solve for the $c_i$
given the experimental data.  For each baryon mass, we assume that the
central value corresponds to the midpoint of the mass range quoted in
the {\it Review of Particle Properties}~\cite{RPP}; we take the one
standard deviation error as half of the stated range.  To determine
the off-diagonal mass matrix elements, we use the mixing angles
extracted from the analysis of strong decays given in
Ref.~\cite{CGKM}, $\theta_{N1}=0.61\pm 0.09$ and $\theta_{N3}= 3.04
\pm 0.15$.  These values are consistent with those obtained in
\cite{CC} from radiative decays.  Solving for the operator
coefficients, we obtain the values shown in Table~\ref{csolve}.

\begin{table}[htb]
\caption{Operator coefficients in GeV, assuming the complete set of
Table~\protect\ref{matel}.} 
\label{csolve}
\newcommand{\m}{\hphantom{$-$}}
\newcommand{\cc}[1]{\multicolumn{1}{c}{#1}}
\renewcommand{\tabcolsep}{0.64pc} 
\renewcommand{\arraystretch}{1.2} 
\begin{tabular}{@{}lllllllll} \hline
$c_1$ & $c_2$ & $c_3$ & $c_4$ & $c_5$ & $c_6$ & $c_7$ & $c_8$ &
$c_9$ \\ \hline
$+0.470$ & $-0.036$ & $+0.369$ & $+0.089$ & $+0.087$ & $+0.418$ &
$+0.040$ & $+0.048$ & $+0.012$ \\
$\pm 0.017$ & $\pm 0.041$ & $\pm 0.208$ & $\pm 0.203$ & $\pm 0.157$
& $\pm 0.085$ & $\pm 0.074$ & $\pm 0.172$ & $\pm 0.673$ \\\hline
\end{tabular}\\[2pt]
\end{table}

        Naively, one expects the $c_i$ to be of comparable size. Using
the value of $c_1$ as a point of comparison, it is clear that there
are no operators with anomalously large coefficients.  Thus, we find
no conflict with the naive $1/N_c$ power counting rules.  However,
only three operators of the nine, ${\cal O}_1$, ${\cal O}_3$, 
and ${\cal O}_6$, have coefficients that are statistically distinguishable
from zero!   A fit including those three operators alone is shown in
Table~\ref{3param}, and has a $\chi^2$ per degree of freedom is $1.87$.
Fits involving other operator combinations are studied in 
Refs.~\cite{CCGL1,CCGL2}. Clearly, large $N_c$ power counting is not 
sufficient by itself to explain the $\ell=1$ baryon masses---the 
underlying dynamics plays a crucial role. 
\begin{table}[htb]
\caption{Three parameter fit using operators ${\cal O}_1$, ${\cal
O}_3$, and ${\cal O}_6$, giving $\chi^2/{\rm d.o.f.}=$
$11.19/6=1.87$. Masses are given in MeV, angles in radians.}
\label{3param}
\newcommand{\m}{\hphantom{$-$}}
\newcommand{\cc}[1]{\multicolumn{1}{c}{#1}}
\renewcommand{\tabcolsep}{1.3pc} 
\renewcommand{\arraystretch}{1.2} 
\begin{tabular}{@{}lllllll}  \hline
               & Fit       & Exp.\       && & Fit       & Exp.\      
\\ \hline
$\Delta(1700)$ & $1683$ & $1720\pm 50$ &&$N(1520)$ & $1530$ & $1523\pm
8$ \\
$\Delta(1620)$ & $1683$ & $1645\pm 30$ &&$N(1535)$ & $1503$ & $1538\pm
18$\\
$ N(1675) $ & $1673$ & $1678\pm 8$ && $\theta_{N1}$ & $0.45$ &
$0.61\pm 0.09$ \\
$ N(1700) $ & $1725$ & $1700\pm 50$ && $\theta_{N3}$ &$3.04$& $3.04\pm
0.15$ \\
$N(1650)$   &     $1663$ & $1660\pm 20$ &&  & & \\\hline
\end{tabular}\\[2pt]
Parameters (GeV): $c_1=0.461\pm 0.005$,
$c_3=0.360\pm 0.059$, $c_6=0.453\pm 0.030$
\end{table}

\section{Interpretation and Conclusions} \label{sec:concl}

We will now show that the preference in Table~2 for two nontrivial 
operators, $\frac{1}{N_c} \ell^{(2)} g \,G_c$ and $\frac{1}{N_c} S_c^2$, 
can be  understood in a constituent quark model with a single pseudoscalar 
meson exchange, up to corrections of order $1/N_c^2$.  The argument goes 
as follows:

The pion couples to the quark axial-vector current so that the 
$\overline{q}q\pi$ coupling introduces the spin-flavor structure
$\sigma^i \tau^a$ on a given quark line. In addition, pion exchange 
respects the large $N_c$ counting rules given in Section~2.  A single 
pion exchange between the excited quark and a core quark is mapped
to the operators $g^{ia} G_c^{ja} \ell^{(2)}_{ij}$ and  
$g^{ia} G_c^{ia}$, while pion exchange between two core quarks yields 
$G^{ia}_c G^{ia}_c$. These exhaust the possible two-body operators that 
have the desired spin-flavor structure. The first operator is one of the
 two in our preferred set. The third operator may be rewritten 
\begin{equation}
2 G^{ia}_c G^{ia}_c = C_2 \cdot 1 - \frac{1}{2}T^a_c T^a_c 
- \frac{1}{2} S_c^2
\label{eq:cas}
\end{equation}
where $C_2$ is the SU(4) quadratic Casimir for the totally symmetric core
representation (the {\bf 10}  of SU(4) for $N_c=3$).  Since the core 
wavefunction involves two spin and two flavor degrees of freedom, and is 
totally symmetric, it is straightforward to show that  $T_c^2=S_c^2$.  Then
Eq.~(\ref{eq:cas}) implies that one may exchange $G^{ia}_c G^{ia}_c$ in 
favor of the identity operator and $S_c^2$, the second of the two operators 
suggested by our fits.

The remaining operator, $g^{ia} G_c^{ia}$, is peculiar in that
its matrix element between two nonstrange, mixed symmetry states is given 
by~\cite{CCGL1} 
\begin{equation}
\frac{1}{N_c} \langle gG \rangle = - \frac{N_c+1}{16 N_c} +
\delta_{S,I}\frac{I(I+1)}{2N_c^2}  \,\,\, ,
\end{equation}
which differs from the identity only at order $1/N_c^2$.  Thus to order 
$1/N_c$, one may make the
replacements
\begin{equation}
\{1 \mbox{ , } g^{ia} G_c^{ja} \ell^{(2)}_{ij}  \mbox{ , }
g^{ia} G_c^{ia} \mbox{ , } G^{ia}_c G^{ia}_c \} \Rightarrow
\{1  \mbox{ , }  g^{ia} G_c^{ja} \ell^{(2)}_{ij} \mbox{ , }
S_c^2\} \,\,\, .
\end{equation}
We conclude that the operator set suggested by the data may
be understood in terms of single pion exchange between quark
lines.  This is consistent with the interpretation of
the mass spectrum advocated by Glozman and Riska~\cite{gloz}.
Other simple models, such as single gluon exchange, do not
directly select the operators suggested by our analysis and
may require others that are disfavored by the data.

\end{document}